\documentclass[english,russian,12pt,twoside]{article}
\usepackage{amssymb,amsmath}
\usepackage[russian]{babel}
\usepackage[cp866]{inputenc}
\begin{document}

\vspace{5mm} \textbf{MSQ} 46S10, 53C80

\begin{center} {{\bf   BIQUATERNIONS ALGEBRA AND ITS APPLICATIONS}\\{\bf BY SOLVING OF SOME THEORETICAL PHYSICS EQUATIONS }}
\end{center}
\begin{center}
{\bf L.A.Alexeyeva }
\end{center}
\vspace{2mm}

\centerline{\textit{ Mathematics Institute, Almaty, Kazakhstan
}}
\centerline{alexeeva@math.kz }\vspace{3mm}
Biquaternions algebra is complex expansion of quaternions algebra, which was elaborated by W.R.Hamilton
in  the middle  of XIX century  [1]. It is very useful and very suitable apparatus for description of many physical
processes.     In last decennial these algebras become be actively used in research  of different problems
of electrodynamics [2-4], quantum mechanics [5-9], fields theory [10-13] and others.

This algebra gives possibility to write many well-known systems of differential equations  in  theoretical
 physics in form of one biquaternionic equation in \emph{bigradiental} form. Concept \emph{bigradient}
  is theoretical generalization of
\emph{gradient} concept, differential operator  which charac\-teri\-zes a direction and intensity of exchange of scalar  field.
By analogy bigradient characterizes a direction and intensity of exchange of  biquaternionic fields which are used for description
of complex scalar-vector physical fields.

Here we elaborate the differential algebra of biquaternions on Minkowski space for
construc\-tion of generalize solutions of bigradiental biquaternionic equations, which are equivalent to the hyperbolic
systems of differential equations of 8 order. We consider biquaternionic wave equation (\emph{biwave} equation) and obtain on the base of Fourie transform
its fundamental and generalized solutions.  Shock waves are researched and conditions on their fronts are determined.
Invariance of biwave equation for the groups of orthogonal transforms,  Lorentz and Poincar\'{e} transforms are explored and relativistic
formulas are constructed.The new scalar equations for
potentials of bigradiental fields (\emph{KGFSh-equation})) is obtained,
which contains Klein-Gordon-Fokk and Shr\'{o}dinger operators, and its generalized solutions are  defined.
The biquaternionic representation of Maxwell and Dirac equations are considered,
their fundamental and generalized solutions are obtained, which describe nonstationary,
harmonic and static scalar-vector electro\-mag\-netic fields, spinors and spinors fields.

Properties of superposition of bigradiental operators, which is very easy calculated in biqua\-ternions algebra,
 and methods of genera\-lized functions theory
make the process of solutions construction simple and fine, as  you'll see hereinafter.

\bigskip

\centerline{ \textbf{1. Biquaternions algebra}}\vspace{2mm}

 We introduce some notions and indications, which will be used here.
Let $e_1 ,e_2 ,e_3 $ are basis vectors of cartesian coordinate system in $R^3$ , $e_0=1$;
 $F$ is three dimensional vector with complex components:
$F = F_1 e_1  + F_2 e_2  + F_3 e_3 $,  $F_j,f\in \mathcal{C}$ are complex numbers;
 $\varepsilon _{jkl} $ is Leavy-Chivitta pseudotensor, $\delta _{jk} $ is Kronecker symbol.

We consider the linear space of hypercomplex numbers (\emph{biquaternions})
 $B=\{ {\bf F} = f + F\} $ :
$$a{\bf F} + b{\bf G}
= a(f + F) + b(g + G) = (af + bg) + (aF + bG),\quad \forall a,b\in \mathcal{C},$$
with operation of quaternionic multiplication [1]:
 $${\bf F} \circ
{\bf G} = (f + F) \circ (g + G) = (fg - (F,G)) + (fG + gF + [F,G]).
\eqno(1) $$
Here and hereinafter we designate as $(F,G) = \sum\limits_{j = 1}^3 {F_j G_j
} $ the scalar product of vectors $F$ and $G$; their vector product is $\left[ {F,G}
\right] ={\sum\limits_{j,k,l = 1}^3 \varepsilon _{jkl} F_j G_k } e_l
$ .

Biquaternions algebra is associative:
\[
{\bf F} \circ{\bf G} \circ {\bf  H} = \left( {{\bf F} \circ {\bf G}}
\right) \circ {\bf  H} = {\bf F} \circ \left( {\bf G} \circ
{\bf  H} \right),\eqno(2)
\]
but not commutative:
\[
<{\bf F} ,{\bf G}>={\bf F} \circ{\bf G} - {\bf G} \circ {\bf F}=2[F,G].\eqno(3)
\]
It may be proved on the base of these properties of Leavy-Chivitta pseudotensor and basic elements:
$$\varepsilon _{jkl}  = \varepsilon _{ljk}  = \varepsilon _{klj},
\quad \varepsilon _{jkl}  = -\varepsilon _{jlk}= -\varepsilon _{kjl};\,\,
\varepsilon _{jkl} \varepsilon _{mnl}  = \delta _{jm} \delta
_{kn}  - \delta _{jn} \delta _{km} ,   $$
\[e_0
\circ e_0  = e_0 ,\quad e_0  \circ e_j  = e_j ,\quad e_j  \circ
e_k  =  - \delta _{jk}  + \varepsilon _{jkl} e_l ,\]
 \[ \left(
{e_j  \circ e_k } \right) \circ e_m  =  - \varepsilon _{jkm}  -
\delta _{jk} e_m  - \delta _{km} e_j  + \delta _{mj} e_k  = e_j
\circ \left( {e_k  \circ e_m } \right),\quad   j,k,l,m,n = 1,2,3, \]
( over similar indexes in product there are summation from 1 to 3,   like to tensor convolution).

From the properties  of commutator (3) it follows\\
\emph{Jacobi identity}
$$< <{\bf F},{\bf G}>,{\bf  H}>  +< <{\bf H},{\bf F}>,{\bf G}> +< <{\bf G},{\bf H}>,{\bf F}> =$$
$$=4[\, [F,G], H]  +4[ \,[H,F],G]+4[ \,[G,H], F] =$$
$$=-4(F(H,G)-G(H,F)+H(G,F)-F(G,H)+G(F,H)-H(F,G))=0.$$
 I.e. biquaternions algebra  is \emph{Lie algebra}.

 From (3) we see, that the product of two biquaternions is commutative if one of them is a scalar or
 if their vector parts are parallel.\bigskip

\textbf{ \emph{Some definitions}.}\vspace{2mm}

Bq. ${\bf
F}^ -   = f - F$ is called  \emph{mutual }for ${\bf F} = f + F$.

Bq. ${\bf \bar F} = \bar f + \bar F$ is named  \emph{complex-conjugate } to ${\bf F}$ ( overline marks the complex conjugate number).
 If ${\bf F} \circ {\bf \bar F} = {\bf \bar F} \circ {\bf F} = 1$, then we name ${\bf F}$
\emph{unitary.}

${\bf F}^*  = \bar f -
\bar F$ is termed  \emph{conjugate } to ${\bf F}.\quad$
 If ${\bf F}^*  = {\bf
F}$, then it is  \emph{selfconjugated.}\\
Selfconjugated biquaternion has the form: ${\bf F} = f + iF ,  $ where $f$
and $F$ are real.\vspace{2mm}

\emph{Scalar product }of
${\bf F}_1 ,{\bf F}_2 $ is the bilinear operation
$\left( {{\bf F}_1 ,{\bf F}_2 } \right) = f_1 f_2  + \left( {F_1 ,F_2 }
\right).$\vspace{2mm}

\emph{ Norm} of  ${\bf F}$  is nonnegative scalar value
\[ \left\| {\bf F} \right\| = \sqrt {\left(
{{\bf F},{\bf \bar F}} \right)}  = \sqrt {f \cdot \bar f + \left( {F,\bar F} \right)}  = \sqrt
{\left| f \right|^2  + \left\| F \right\|^2 }.\eqno(4)\]

If ${\bf F}$   is real biquaternion (\emph{quaternion}) then ${\bf F}^*  = {\bf F}^ -  $  and
$$\left\| {\bf F} \right\|^2  =
{\bf F}^*  \circ {\bf F} = {\bf F} \circ {\bf F}^* .\eqno (5) $$

\emph{Pseudonorm } of ${\bf F}$   is the value
\[
\langle\langle{\bf F}\rangle\rangle  = \sqrt {f \cdot \bar f - \left( {F,\bar F} \right)}  = \sqrt
{\left| f \right|^2  - \left\| F \right\|^2 },\quad Re \langle\langle{\bf F} \rangle\rangle\geq 0.\eqno (6) \]
 It's easy to see, if ${\bf F}$ is   selfconjugated then ${\bf \bar F}={\bf F}^-$ and
$$  {\bf \bar F} \circ {\bf F} = {\bf F}
\circ {\bf \bar F}=\langle\langle {\bf F}\rangle\rangle ^2  . \eqno (7)$$

 If ${\bf F} \circ {\bf { G}} = 1$
  then ${\bf G}$   is \emph{right inverse}  for
${\bf F}$   and it is termed as ${\bf {G}} = {\bf F}^{ - 1} $, thereafter ${\bf F}$   is \emph{left unverse} for ${\bf G}$
and it is termed as  $^{ - 1} {\bf { G}}. $

Simply the   equalities are proved: $$\left( {{\bf F} \circ {\bf G}} \right)^*
= {\bf G}^*  \circ {\bf F}^* ,\quad \left( {{\bf F} \circ {\bf G}}
\right)^{ - 1}  = {\bf G}^{ - 1}  \circ {\bf F}^{ - 1}.\eqno (8) $$

Also simply the next lemma and theorem are proved.\vspace{2mm}

\textbf{Lemma 1.1. }\emph{If}$\left(
{{\bf F},{\bf F}} \right) \ne 0$, \emph{then both inverse biquaternions exist and they are equal}:
$${\bf F}^{ - 1}  = ^{ - 1} {\bf F} =
{\bf F}^ -  /({\bf F},{\bf F}).  \eqno(9)$$
\emph{If}$\left(
{{\bf F},{\bf F}} \right) = 0$ then \emph{ inverse Bq. does not exist}.\vspace{2mm}

\textbf{Theorem 1.1.} \emph{By known} ${\bf F}$  and  ${\bf B}$   \emph{biquaternionic linear (bilinear)
 equations of a kind}
 \[{\bf F} \circ {\bf G} = {\bf
{B}} \quad\ and \quad {\bf G} \circ {\bf F} = {\bf {
B}} \]
\emph{have unique solution} ${\bf G} = {\bf F}^{ -
1}  \circ {\bf { B}}$  \emph{and}   ${\bf G} = {\bf { B}} \circ
{\bf F}^{ - 1} $ \emph{appropriately, if and only if} $\left( {{\bf F},{\bf F}}
\right) \ne 0.$\vspace{2mm}

Proof of this theorem follows from the lemma  1.1.

If $\left( {{\bf F},{\bf F}} \right) = 0$,
then existence of the decision depends on  $\bf B$ (see [14]).\vspace{2mm}

We introduce here the very useful Bq.
$${\bf{\Xi}}=W+iP=\frac{1}{2}\,\textbf{F}\circ
\textbf{F}^{\ast}=(|f|^2+\|F\|^2)/2+i\left(\texttt{Im}\,
(\overline{f}F)+ [\texttt{Re}\,F,\texttt{Im}\,F]\right).
$$
$$\langle\langle{\bf{\Xi}}\rangle\rangle^2=W^2-\|P\|^2={\bf{\Xi}}\circ{\bf{\Xi}}^{-}.$$
Here $W$ и $P$ are real values.
It  describes the energy-impulse density of scalar-vector field in  problems of theoretical physics.

 \bigskip

\centerline{\textbf{2. Lorentz and Poincar\'{e} transformations on Minkowski space
} }\vspace{2mm}

Let consider biquaternions on Minkowski space ${M} =
\left\{ {(\tau ,x):\tau  \in R^1 ,x \in R^3 } \right\}$ and groups of linear transformations on it.
We quoternize $M$ by using complex-conjugate Bqs.: $$ {\bf Z} = \tau  +
ix,\quad {\bf \bar Z} = {\bf Z}^-=\tau  - ix ,\,\quad\tau\in
R^1,\,\,x\in R^3.$$ They are selfconjugated: $ {\rm }{\bf Z} = {\bf
Z}^* ,\,\,{\bf \bar Z} = {\bf \bar Z}^*$,
 and have equal norms and pseudonorms:
$$\left\| {\bf Z} \right\|^2 =\left\| {\bf \bar Z} \right\|^2  = \tau^2+\|x\|^2=({\bf Z},{\bf \bar Z}),\quad
\langle\langle {\bf Z} \rangle\rangle ^2=\langle\langle {\bf \bar
Z} \rangle\rangle ^2 = \tau^2-\|x\|^2={\bf Z}\circ{\bf \bar
Z}\quad \eqno(10)$$ and
$${\bf Z}^{ - 1}  = {\bf \bar
Z}/ \langle\langle {\bf Z} \rangle\rangle ^2 ,\quad\quad{\bf \bar
Z}^{ - 1} = {\bf Z}/\langle\langle {\bf Z} \rangle\rangle ^2 .$$
Hence on \emph{light cone }($|\tau|=\|x\|$) inverse Bq.
for ${\bf Z} ,\,\,{\bf \bar Z}$ does not exist.\vspace{2mm}

\textbf{\emph{Orthogonal transformation.}} Let consider conjugate Bqs.:
$$ {\bf U}(\varphi,e) = \cos \varphi  + e\sin \varphi ,\,\, {\bf
U}^{*}={\bf U}^{-}  = \cos \varphi  - e\sin \varphi ,\quad\left\|
e \right\| = 1, \quad \varphi\in R^1.$$
It's easy to see:$$\parallel
\bf{U}\parallel=\parallel \bf{U}^*\parallel=1,\,\,
\bf{U}\circ\textbf{U}^*=\bf{U}^-\circ\bf{U}=1.\eqno (11)$$
\vspace{2mm}

\textbf{ Lemma 2.1.} \emph{Conjugate biquaternions} $ {\bf
U}(\varphi,e), {\bf U}^{*}(\varphi,e)  $,
 \emph{$\varphi\in R^1$, define the group of orthogonal transformation on  М},
\emph{which are orthogonal on vector part of Bq.} \textbf{Z} : $$ {\bf
Z}^{\bf '}  = {\bf U} \circ {\bf Z} \circ {\bf U}^{*} ,\,\,\, {\bf
Z} = {\bf U}^{*}  \circ {\bf Z}^{\bf '}  \circ {\bf U}.$$ \emph{
This transformation is the rotation of space $R^3$ around vector $e$ through angle  $2\varphi$ }.\vspace{2mm}

\textbf{Proof. } By calculating lemmas formulae we get conservation of scalar part
and specified rotating of vector part:
\[
\tau ' =\tau,\quad
 x' =e(e,x) +
(x - e(e,x))cos 2\varphi +  [e,x]sin 2\varphi .
\]
Under (10)-(11) the pseudonorm
$
\langle\langle {{\bf Z'}} \rangle\rangle ^{\bf 2}  = {\bf U} \circ
{\bf Z} \circ {\bf U}^{*} \circ {\bf \bf U} \circ {\bf \bar Z }
\circ {\bf  U ^{*}} = \langle\langle {\bf Z} \rangle\rangle ^{\bf
2}$.
Since $\tau=\tau'$, the norm of vector $Z$ hold true: $\|Z\|=\|Z'\|$.\vspace{2mm}

 Let consider superposition of two orthogonal transformation:
\[
\displaylines{ {\bf U}_1  \circ {\bf U}_{\bf 2}   = {\bf U}_3  =
u_3  + U_3   = (\cos \varphi _1  + e_1 \sin \varphi _1 ) \circ
(\cos \varphi _2  + e_2 \sin \varphi _2 ) =  \cr
   = \cos \varphi _1 \cos \varphi _2  - (e_1 ,e_2 )\sin \varphi _1 \sin \varphi _2  + e_2
    \sin \varphi _2 \cos \varphi _1  + e_1 \sin \varphi _1 \cos \varphi _2  +  \cr+[e_1 ,e_2 ]\sin \varphi _2
     \sin \varphi _2 .\cr}
\]
Since
\[
   {\bf U}_3^{ - 1}  = \left( {{\bf U}_1  \circ {\bf U}_{\bf 2} } \right)^{ - 1}  = {\bf U}_2^{ - 1}  \circ {\bf U}_1^{ - 1}  = {\bf U}_2^ - \circ {\bf U}_1^ -
   ,\]
\[  {\bf U}_3^*  = \left( {{\bf U}_1  \circ {\bf U}_{\bf 2} } \right)^*  = {\bf U}_2^*  \circ {\bf U}_1^*  = {\bf U}_2^ -   \circ {\bf U}_1^ -   = {\bf U}_3^
  -,
\]
we get:
\[{\bf U}_3^{}  \circ {\bf U}_3^ -   = \left( {{\bf U}_1  \circ
{\bf U}_{\bf 2} } \right) \circ \left( {{\bf U}_1  \circ {\bf
U}_{\bf 2} } \right)^ -   = {\bf U}_1  \circ {\bf U}_{\bf 2} \circ
{\bf U}_2^ -   \circ {\bf U}_1^ -   = 1
\]
Hence ${\bf U}_3 $ is also orthogonal transformation:
\[
{\bf U}_3  =  \cos \varphi _3  + e_3 \sin \varphi _3 , \quad
\varphi _3= arc (\cos u_3), \quad e_3=\frac{U_3}{\|U_3\|.}
\]
Lemma has been proved.
 \vspace{2mm}

\emph{\textbf{Lorentz transformations.} }Let consider mutual selfconjugated Bgs.
$$ {\bf L} (\theta,e)=
ch\,\theta  + ie\,sh\,\theta ,\,\,{\bf  L^{-}}=ch\,\theta  -
ie\,sh\,\theta ,\quad \theta\in R^1,\,\, \left\| e \right\| =1
$$  (here hyperbolic sine and cosine are used).
It's easy to see that they are unitary: $${\bf L}\circ{\bf
L^{-}}=ch\theta ^2-sh\theta ^2=1 .\eqno (12)$$
Next lemmas are proved by use of simply calculations
[14].\vspace{1.5mm}

   \textbf{ Lemma 2.2. } \emph{Lorentz transformation has such biquaternionic representation}:   $$
{\bf Z}' = {\bf L} \circ {\bf Z} \circ {\bf L},\quad {\bf Z} =
{\bf  L ^{-}}
  \circ {\bf Z}' \circ {\bf L ^{-}},\quad\langle\langle {{\bf Z'}} \rangle\rangle ^{\bf 2}  =
\langle\langle {\bf Z} \rangle\rangle ^{\bf 2}.\eqno (13)$$
Also easy to show that pseudonorm conserves on the associativity and unitarians (12):
 \[
\langle\langle {{\bf Z'}} \rangle\rangle ^{\bf 2}  = {\bf L} \circ
{\bf Z} \circ {\bf L} \circ {\bf \bf L^{-}} \circ {\bf \bar Z }
\circ {\bf  L^{-}} = \langle\langle {\bf Z} \rangle\rangle ^{\bf
2}.
\]
If to enter designations: $$ ch2\theta  =
(1 - v^2 )^{-1/2},\quad
 sh2\theta  = v (1 - v^2 )^{-1/2},\quad\left| v
\right| < 1 ,$$
After calculating we get known formulas for scalar and vector parts of ${\bf Z}'$ and ${\bf Z}$  [13]:\\\emph{relativistic formulas }
\[
\tau ' = \frac{{\tau  + v(e,x)}}{{\sqrt {1 - v^2 } }},\quad x' =
(x - e(e,x)) + e\frac{{(e,x) + v\tau }}{{\sqrt {1 - v^2 } }},
\]
\[
\tau  = \frac{{\tau ' - v(e,x)}}{{\sqrt {1 - v^2 } }},\quad x =
(x' - e(e,x')) + e\frac{{(e,x') - v\tau '}}{{\sqrt {1 - v^2 } }},
\]
It corresponds to motion of coordinates system $\{X_1,X_2,X_3\}$ in direction of vector \emph{е }with velocity \emph{v}.

Superposition of two Lorentz transformations with equal $e_j$ possess with group  properties:
\[
\displaylines{
  {\bf L}_1  \circ {\bf L}_{\bf 2}  = {\bf L}_3  = l_3  + L_3  = (ch\varphi _1  +
   iesh\theta _1 ) \circ (ch\theta _2  + iesh\theta _2 ) =  \cr
   = (ch\theta _1 ch\theta _2  + sh\theta _1 sh\theta _2 ) + ie(sh\theta _2 ch\theta _1
    + sh\theta _1 ch\theta _2 ) =  \cr=ch(\theta _1  + \theta _2 ) + esh(\theta _1  + \theta _2 ) \cr}
\]
But superposition of two Lorentz transformations with nonparallel $e_1$ and
$e_2$ does not constitute the Lorentz transformation:
\[
\displaylines{
  {\bf L}_1  \circ {\bf L}_{\bf 2}  = {\bf L}_3  = l_3  + L_3  = (ch\theta _1  + ie_1 sh\theta _1 ) \circ (ch\theta _2  + ie_2 sh\theta _2 ) =  \cr
   = (ch\theta _1 ch\theta _2  + (e_1 ,e_2 )sh\theta _1 sh\theta _2 ) + i(e_2 sh\theta _2 ch\theta _1  + e_1 sh\theta _1 ch\theta _2 ) - [e_1 ,e_2 ]sh\theta _2 sh\theta _1  \cr}
\]
What see, this Bq. contains real value in vector part. \vspace{2mm}

\emph{\textbf{Poincar\'{e} transformation. }} Note that  $${\bf
L}(\theta,e)={\bf U}(-i\theta,e).$$ Using superpositions of  these two transformations we get general form
of the transformation, which may be named  \emph{Poincar\'{e} transformation}.\vspace{2mm}

\textbf{Definition.} Poincar\'{e} transformation on \emph{M } is the linear transformation of the type:
\[
{\bf Z}^{\bf '}  = {\bf P} \circ {\bf Z} \circ {\bf P^*} ,\quad\quad
{\bf Z} ={{\bf P^-}}  \circ {\bf Z}^{\bf '}  \circ {{\bf
P^*}^-},
\]
$${\bf P} = {\bf U} \circ {\bf L} = cos(\varphi  +i\theta
) + esin(\varphi  +i\theta ), \quad\quad {\bf P}^*  = {\bf L}^* \circ
{\bf U}^* =  cos(\varphi -i\theta ) -esin(\varphi  -i\theta ),$$
which conserves the pseudonorm:
$\langle\langle {\bf{Z}}\rangle\rangle=\langle\langle\textbf{Z}'\rangle\rangle.$

It's easy to prove because ${\bf  P} \circ {\bf P}^-  ={\bf P^*} \circ {{\bf
P^*}^-} = 1 $.
By such transformations light cone $( \tau=\|x\|)$ is invariant set as
 $\langle\langle\textbf{Z}\rangle\rangle=\langle\langle\textbf{Z}^-\rangle\rangle=0.$

 Superposition of two Poincar\'{e} transformations with equal $e_j$  also possess with group  properties.
 It's easy to show:
 $$ \textbf{P}_1\circ \textbf{P}_2=\textbf{L}_1\circ \textbf{U}_1\circ  \textbf{L}_2\circ  \textbf{U}_2=
 \textbf{L}_1\circ  \textbf{L}_2\circ  \textbf{U}_1\circ  \textbf{U}_2=\textbf{L}_3\circ \textbf{U}_3=\textbf{P}_3.$$
But   superposition of two Poincar\'{e} transformations with nonequal $e_j$ doesn't possess such properties.

\bigskip

\centerline{\textbf{3. The space of generalized biquaternions ${{\texttt{B}'}}{(M)}$}} \vspace{3mm}

We will consider on $\emph{M}$ the functional  space of Bqs.  $\texttt{B}(\emph{M}) = \{
{\bf  F} = f(\tau,x ) + F(\tau,x )\}$, where $f$ is complex function and $F$ is three dimensional vector-function
with complex components $F_j$, $j=1,2,3$.
The partial derivative from Bq. over  $\tau$ or $x_j$
 we designate so:$$\partial_\tau {\bf  F}=\partial_\tau f + \partial_\tau F,\quad
 \partial_j {\bf  F}=\frac{\partial f}{\partial x_j} + \frac{\partial F}{{\partial x_j}}\,,\quad j=1,2,3.$$

Let introduce two biquaternions spaces,  \emph{basic} one is
 $$\texttt{B}(M)=\{{\bf \Phi}=\varphi(\tau,x)+\Phi(\tau,x) \}, \quad\varphi\in  D(R^4),\,\Phi_j\in  D(R^4),\,j=1,2,3,$$
where $D(R^4)$ is the space of  finite infinitely differentiable functions on  $R^4$,
and conjugate space ${\texttt{B}'} (M)=\{{\bf \hat F}=\hat{f}+\hat{ F}\}$ of linear continues functionals on
${\texttt{B}(M)}$  :
$$({\bf \hat F},{\bf\Phi })=(\hat{f},\varphi)+\sum_{j=1}^{3}(\hat{F}_j,\Phi_j ),
\quad \forall {\bf  \Phi }\in {\texttt{B}(M)},
$$
which be named the  space of {\emph{generalized biquaternions}} and we mark such Bq. with cap over it.

Any regular Bq.${\bf  F}$ corresponds to the functional,
which can be presented in integral form:
\[
({\bf{\hat F}},{\bf{\Phi}})=
\int\limits_{R^4 } {\left({\bf{F}}(\tau,x),{\bf{\Phi}}(\tau,x)\right) d\tau dx_1 dx_2 dx_3},\quad \forall {\bf \Phi }\in \texttt{B}(M).
\]

Bq. is \emph{singular} if its action on $\texttt{B}(M)$ can not be presented in such form.
Example of singular Bq. is  singular functions from $D'(R^4)$ because $D'(R^4)\subset {\texttt{B}'} (M)$.
Using them the more complex biquaternions may be constructed.

In particular, in the mathematical physics problems the next singular functions are often used as \emph{simple layers}.
  There generalization on  $\hat{ B}(M)$ are generalize Bqs.
${\bf F}\delta _S$, which define the functionals of the type:
\[
({\bf F}\delta _S,{\bf \Phi }) = \int\limits_{S} {\left( {{\bf F}(\tau ,x),{\bf \Phi }(\tau ,x)}
\right)} dS, \quad \forall {\bf \Phi }\in \texttt{B}(M).
\]
Here is the surface integral on a surface  $S \subset R^4$, which dimension may be equal 1,2,3.
Defined and integrable on \emph{S} Bq. ${\bf F}$
we name the \emph{ density of simple layer  } as is customary. \vspace{2mm}

\emph{\textbf{Differentiation}}. Analogically as in the theory of generalized function [15]
using the definition of the partial derivatives of  generalized Bq.:
$$({\partial_{j}\bf \hat F},{\bf \Phi })=-({\bf \hat F},\partial_{j}{\bf  \Phi })\quad\textrm{ for}
\quad \forall {\bf  \Phi }\in {\texttt{B}(M)}
$$
 the derivatives of singular Bqs. and the derivatives of more higher order can be constructed.

From the properties of differentiation of regular functions with finite gaps
 on some 3-dimensional surface $S\subset R^4$ there is
$${\partial_{j}\bf \hat F}={\partial_{j}\bf  F}+n_{j}[{\bf F}]_{S}\delta _S,\quad j=\tau,1,2,3 \eqno (14)$$
Here ${\partial_{j}\bf  F}$  is classic derivative, $n=\{n_\tau,n_1,n_2,n_3\}$ is unit normal to  $S$.
In square brackets the gap of ${\bf F}$ on $S$ stands,
\[\left[ {{\bf F}(\tau ,x)}
\right]_S  = \mathop {\lim }\limits_{\varepsilon  \to  + 0}
\left\{ {{\bf F}(\tau  + \varepsilon n_\tau  ,x + \varepsilon n) -
{\bf F}(\tau  - \varepsilon n_\tau  ,x - \varepsilon n)} \right\}
,\quad (\tau ,x) \in S .\]

  \textbf{Definition. }Bq. {\bf F } is  \emph{generalize solution} of differential equation
 $D(\partial_\tau,\partial_ x){\bf F }= {\bf G}$
 if
\[
({D\hat{\bf F},{\bf \Phi } }) = (\hat{{\bf G}},{\bf \Phi }) \quad\textrm{ for}\,\,
 \forall {\bf  \Phi }\in {\texttt{B}(M)}
\]
(here \emph{D } is linear differential operator). If  $\hat{\bf F}$ is quasi everywhere differentiable regular
function ${\bf F}$ then it is classic solution.

\textbf{\emph{Convolution}} of two biquaternions is the bilinear operation:
\[
{\bf A}(\tau ,x)*{\bf B}(\tau ,x) = a*b - \sum\limits_{i,j,l =
1}^3 {\left( {A_j *B_j } \right)}  + \left( {a*A_j } \right)e_j  +
\left( {b*B_j } \right)e_j  + \varepsilon _{ijl} \left( {A_i *B_j
} \right)e_l,
\]
where usual convolutions  of generalized functions [15] stand in brackets on the right.
It's easy to see, here two operations of biquaternionic
multiplication and convolution are united.

In virtue of properties of convolution differentiation [15] we get the formula for derivative  of Bqs. convolution:
\[\partial_j ({\bf A}*{\bf B})=(\partial_j {\bf A})*{\bf B}={\bf A}*\partial_j {\bf
B},\quad \partial_j=\partial_\tau,\partial _1 ,\partial _2
,\partial _3. \eqno(15)
\]
It gives possibility to use more convenient form by its calculating .
\vspace{2mm}

\emph{\textbf{ Fourier transformation}}.
\emph{Generalized} Fourier transformation (FTr.) of
${\hat{\bf G}}$ is ${\bf \tilde {G}}$ which satisfies the equality:
$$ (\hat{\bf {G}},\bf{\Phi})=({\bf \widetilde{G}},{ \bf\tilde {\Phi}})\quad\textrm{for}\,\,
 \forall {\bf \Phi} \in {\texttt{B}(M)}.$$
Here ${\bf \tilde {\Phi}}$ is classic FTr. of ${\bf {\Phi}}$:
$$\texttt{F}[{\bf \Phi}(\tau ,x)]={\bf \tilde {\Phi}}(\omega, \xi)=\int\limits_{R^4 }
{\bf{{\Phi}}}(\tau,x)exp(i\tau\omega+i(\xi,x))d\tau dx_1dx_2dx_3,\quad {\bf \Phi} \in {\texttt{B}(M)},$$
which always exists in virtue of properties $\texttt{B}(M)$.
If Bq. is regular and has classic  FTr. then it is also generalized FTr.

Also it's easy to prove (similarly in [15]), that
\[
\texttt{F}[{\bf A}(\tau ,x)*{\bf B}(\tau ,x)]={\bf
\widetilde{A}}(\omega ,\xi)\circ{\bf \widetilde{B}}(\omega ,\xi).
\eqno(16)\]

Using properties of FTr. of generalized functions we could get many similar properties of  Fourier transformation
 for biquaternions,   but we don't do this here.
\bigskip

\centerline{\textbf{4. Bigradients.  Poincar\'{e} and Lorentz transformations}}\vspace{2mm}
Let consider special cases of differential operators which are representative for mathematical physics equations.
But we will analyze them on  ${ \texttt{B}'}(M)$.

We introduce operators, named \emph{mutual complex gradients}:
 $$\nabla^ +   = \partial _\tau   + i\nabla
,\quad \nabla^ {-} = \partial _\tau   - i\nabla,\eqno(18)$$
where $\nabla  = grad = (\partial _1 ,\partial _2 ,\partial  _3  )$.
We name them shortly \emph{bigradients}.

In the sense of given definitions their symbols are complex conjugate and selfconjugated: $(\nabla^ - )^*
= \nabla^ -  ,\;(\nabla^ +  )^*  = \nabla^ + . $
Their action is defined in according to biquaternions algebra
 $$
   \nabla^ \pm  {\bf F} = (\partial _\tau   \pm i\nabla ) \circ (f +
   F) = (\partial _\tau  f \mp i\,(\nabla ,F)) \pm\partial _\tau  F \pm
                      i\nabla f \pm i[\nabla ,F]
  $$
  (correspond to signs)  or in conventional record
  $$  \nabla^ \pm  {\bf F} =  (\partial _\tau  f \mp i\,div\,F) \pm \partial _\tau  F \pm igrad\,f
                            \pm i\,rot\,F.$$
It's easy to test that wave operator ( $ \Box$) is presented in the form of superposition
of mutual bigradients:
$$ \Box=\frac{\partial^2}{\partial\tau^2}-\triangle= \nabla^ -   \circ \nabla^ +
= \nabla^ +   \circ \nabla^ -  ,\eqno (19)$$
where $\triangle=\sum\limits_{j=1}\limits^3 \partial_j\partial_j$ is Laplace operator.
Using this property, we can construct  particular solutions of differential biquaternionic equations on
$\texttt{B}'(M)$ of a type:
\[
 \nabla^ \pm  {\bf K}(\tau,x) = {\bf G}(\tau,x), \eqno(20)
\]
which be named \emph{biwave equation}. Solutions of this equations be named
\textit{$\pm$bipotentials } of ${\bf G}$.

By Poincar\'{e} transformations bigradients and biwave equations transform according consecutive affirmations.\vspace{2mm}

\textbf{   Theorem 4.1.} \emph{By Poincar\'{e} transformations }
(${\bf Z} \rightarrow {\bf Z}^\textbf{'}  = {\bf P} \circ {\bf Z} \circ {\bf P}^* $)
 \emph{biwave equation transforms to biwave  equation}:
$${\bf D}'{\bf K'} = {\bf G'},$$
$ {\bf D}=\nabla^{\pm} ={\partial_{ \tau} \pm i\,grad_{x}}$, $ {\bf D}'={{\bf  P}^-  \circ \bf{D} \circ
{\bf P}} = {\partial_{ \tau '} \pm i\,grad_{x'}} $,

(\emph{relativistic formulas})
 $$ {\bf K}^{\bf '}  = {\bf  P}^- \circ {\bf K} \circ
{\bf P} ,\quad\quad {\bf G}^{\bf '}  = {\bf  P}^- \circ {\bf G}
\circ {\bf P}.$$

\vspace {2mm}

\textbf{Proof.} See [14] and equalities
\[
\textbf{D}'{\bf K}' = \left( {{\bf  P}^-  \circ \textbf{D} \circ
{\bf P}} \right)\circ\left( {{\bf  P}^-  \circ {\bf K} \circ {\bf
P}} \right) = {\bf  P}^-  \circ \textbf{D} \circ {\bf K} \circ
{\bf P} = {\bf  P}^- \circ {\bf G} \circ {\bf P} = {\bf G'}.
\]

Lets go to solving biwave equations.
\bigskip

\centerline{\textbf{5. Generalized solutions of biwave equations. Shock waves}}\vspace {2mm}

\textbf{Theorem  5.1.} \textit{Generalized solution of biwave equation
} (20) \emph{can be presented in the form}:
$$\hat{{\bf K}} = \nabla^ \mp  \hat{{\bf G}} * \psi+{{\bf K_0}}.
\eqno (21) $$
\textit{where }  $\psi (\tau,x)$ is \emph{ simple layer on light cone} $\tau  = \left\| x \right\|$:
\[\psi  = (4\pi \left\| x \right\|)^{
- 1} \delta (\tau  - \left\| x \right\|), \eqno (22)
\]
\textit{ which is the fundamental solution of wave equation}:
$$\Box\psi =\delta (\tau) \delta(x).\eqno(23)$$
\emph{Bq.} ${\bf K_0}(\tau ,x)$  \textit{is a solution of homogeneous biwave equation} (\emph{by }$\hat{{\bf G}}=0)$.
\emph{It can be presented as}
\[
 {{\bf K_0}} = \nabla^ \mp  \{{\bf
G_0}* \psi_0(\tau ,x)\},\eqno (25)
\]
\textit{where } $\psi_0 (\tau ,x)$ \textit{ is the a solution of
homogeneous wave equation}:
$$\Box\psi_0=0,\eqno (26)$$
\[
\psi_0  (\tau ,x) =  \int\limits_{R^3 } {\phi
(\xi )} \exp \left( {i\left( {(\xi ,x) \pm \left\| \xi
\right\|\tau } \right)} \right)dV(\xi ) ,\quad\forall \phi (\xi ) \in L_1 (R^3 ),  \eqno (27 )
\]
${{\bf
G_0}(\tau ,x)}$\textit{ is arbitrary Bq., capable of this convolution with $\psi_0$,
or it can be presented as a sum of such type solutions.}

\vspace{2mm}

\textbf{Proof. }Using associativity and properties (15), (23) for first summand in  (21)
we get
\[ \nabla^ \pm  \hat{{\bf K}} =
\nabla^ \pm  \nabla^ \mp \left( \hat{{\bf G}} * \psi  \right) = \Box\left( {\hat{{\bf G}} * }
 \psi \right) = {\hat{{\bf G} }* }  \Box{\psi }  =
\hat{{\bf G}}*\delta (\tau )\delta (x) = \hat{{\bf G}}.
\]
Last equality in virtue of the property of  $\delta$-function [15].
By analogy for the second summand:
$${\nabla^
\pm{\bf K_0}} = \nabla^ \pm\,\nabla^ \mp  \{{\bf C_0}* \psi_0(\tau
,x)\}= \Box \{{\bf C_0}* \psi_0(\tau ,x)\}=\{\Box \psi_0(\tau
,x)\}*{{\bf C_0}}=\textbf{0}.$$

Inversely, if $\bf K_0$ is the solution of homogeneous biwave equation,
 then every its components  is the solution of Eq. (26). Thus and so
 it can be presented in the form (25) or decomposed  in sum of forth such bipotentials for every components
 generated with different solutions of Eq. (26).
In virtue of linearity their sum in  (21) is a solution of Eq. (20).

The form of solutions of wave equation (26), including fundamental one (22), is well known [15].
\vspace {2mm}

\textbf{\emph{Fundamental solution of biwave equation}} we obtain if suppose
$\textbf{G}=\delta(\tau)\delta(x)$. Then by use  formula (21) we have  its presentation:

\[{\bf{\Psi}}(\tau,x)=\partial_\tau\psi\pm i\, grad\,\psi.
\]
It gives possibility to construct solutions of biwave equation  in form of the convolution:
\[
\hat{\textbf{K}} =  \hat{{\bf G}} \ast {\bf{\Psi}} + {\bf {K}}_0
\]
for arbitrary right part of Eq.(20), accepting such convolution.
\vspace{2mm}

\emph{\textbf{Shock waves. }}Let consider generalized solutions
 of biwave equation. Whereas it is hyperbolic, there are solutions which are non differentiable   on
characteristic surfaces ($S$) where
\[ n_\tau ^2  = \left\| n \right\|^2 \eqno (28)\]
$(n_\tau  ,n_1 ,n_2 ,n_3 ) $ is unit normal to $S$ in $M$, $n =
(n_1 ,n_2 ,n_3 ),\quad \left\| n \right\|^2  = n_1^2  + n_2^2  +
n_3^2 . $ This is a cone of characteristic normals of wave equation.

Let define the conditions on the gaps of derivatives  for regular solutions of
biwave equation. By using formula (14) we get:
$$\displaylines{ \nabla ^ +  {\bf \hat K} = \nabla ^ +  {\bf K} +
\left\{ {n_\tau  \left[ k \right]_S  - i\,(n,\left[ K \right]_S )
+ n_\tau  \left[ K \right]_S  + in\left[ k \right]_S  + i[n,\left[
K \right]_S ]} \right\}\delta _S (\tau ,x) =  \cr = \nabla ^ +
{\bf K} + (n_\tau   + in) \circ \left[ K \right]_S \delta _S (\tau
,x) = 0 \cr}$$
Because $\nabla ^ +  {\bf K} = 0$ then
$$ n_\tau  [ k]_S  -
i\,(n,[ K ]_S ) = 0,\quad n_\tau  [ K ]_S  + in [ k]_S  + i[n,[ K]_S ] = 0. \eqno (29)$$

Such solution is  named
\emph{shock wave}. In $R^3$  it has  mobile wave front $S_t$ with normal $(n_1,n_2,n_3)$, which propagates
in virtue (28) with speed
\[ c =  -\frac{ n_\tau }{\left\| n \right\|}=  1. \eqno (30)\]

This result is formulated in this theorem.\vspace{2mm}

\textbf{ Theorem  5.2.} \emph{Shock waves satisfy to the conditions on its fronts
}:
$$\left[ \textbf{K }\right]_S  = i{\bf{m }}\circ\left[ \bf{K }\right]_S, $$
 \emph{where }${\bf{m }}=n/\|n\|$ is \emph{a unit wave vector directed along the speed of its expansion
in} $R^3$.

\textbf{Proof. }If to divide  (29) per $\left\| n \right\|$ subject to (30) , we get
$$ \left[ k \right]_S  =  - i\,({\bf{m }},\left[ K \right]_S
) , \quad\left[ K \right]_S  = i\,{\bf{m }}\left[ k \right]_S  + i\,[{\bf{m }},\left[ K
\right]_S ] \eqno (31)$$
Pass to its biquaternionic record we get the formula of the theorem.

First equation (31) characterizes \emph{longitudinal shock waves}.

If to substitute it
to second equation, we get formula for tangent component of vector \emph{К } to wave front: $$\left[ K
\right]_S  - {\bf{m }}({\bf{m }},\left[ K \right]_S ) = i\,[{\bf{m }},\left[ K \right]_S ].\eqno (32)
)$$
They give the connection between gaps of real and imaginary parts of Bq.
\bigskip

{\textbf{ \emph{Generalized Kirchhoff formula for biwave equation}}}
Now we  solve Cauchy problem for biwave equation. Let \emph{initial condition} is given:
\[
{\bf K}(0,x)={\bf K^0}(x).\eqno (33)
\]
We have to construct the solution of Eq.(20), satisfying to these data.

We use for this \emph{generalized functions method}.
Let introduce regular  functions
 ${\bf \widehat{G}}=H(\tau){\bf G}(\tau,x)$, where $H(\tau)$ is Heaviside function.
If  ${\bf K^0}$ is regular then
  $$\nabla^ \pm  {\bf \widehat{K}} = {\bf \widehat{G}}+\delta(\tau) {\bf {K^0}}(x). \eqno (34)$$
Therefore  the solution is
\[
{\bf K}(\tau,x) = \nabla^ \mp  \{H(\tau){\bf {G}} *
\psi\}+{\bf G}(0,x)
 \mathop *\limits_x \psi+\nabla^ \mp\{{\bf K^0}(x) \mathop *\limits_x \psi\},\eqno (35)
 \]
where sign $\mathop *\limits_x$   implies a  convolution only over  $x$.
For regular ${\bf \widehat{G}}$ it has integral representation:
\[
{4\pi }\hat{{\textbf{K}}}(\tau,x) =- \nabla^ \mp  \left\{\int\limits_{r \le \tau } {{{\bf G}(\tau -
r,y)}}\frac{dV(y)}{r} +\tau^{-1}\int\limits_{r =\tau } {\bf K^0}(y) dS(y)\right\}-\tau^{-1}
\int\limits_{r =\tau } {\bf G}(0,y) dS(y).  \eqno (36)
 \]
Here and further  $ r=\|y-x\|,\,dV(y)=dy_1dy_2dy_3$,  $dS(y)$ is differential of spheres surface $r=\tau$ in point $y$.
This formula is generalization of  well known Kirchhoff formula
for solution of wave equation [15].
 \vspace{3mm}

 \centerline{\textbf{ 6. Biquaternionic form of Maxwell equations and its modification}}\vspace{2mm}

As example of applications  of biwave equation theory let consider Maxwell equation for electromagnetic  field.  It contains
 8 differential Eqs. (2 vectorial and 2 scalar Eqs.).

In space of biquaternions it has the form of biwave equation [2,4,16]:
\[
\nabla^+  {\bf A}+{\bf \Theta }=0 .\eqno(37)
\]
Biquaternions of \emph{intensity}$\textbf{A}$ and \emph{charge - current }${\bf \Theta }$ of  EM-field are
$$\textbf{A}=0+A=\sqrt{\varepsilon}\, E(\tau,x)+i\sqrt{\mu }\,H(\tau,x),\quad
{\bf \Theta } = i\rho(\tau,x) + J(\tau,x),\eqno(38)$$
where  ${E}$ is electric field intensity, ${H}$ is magnetic field intensity;
\emph{a density }$\rho$ and \emph{pulse}  $J$   are denominated with use of density of electric charge
 $\rho ^E$ and electric current density  $j^E$ by formula:
\[
\rho  = \rho ^E/\sqrt{\varepsilon}=\sqrt{\varepsilon}\,div\,E , \quad J = \sqrt {\mu }\, j^E  ,
\]
$ \varepsilon,\mu$ are the constants of electric conduction and
 magnetic conductivity;$\tau=ct$, $t$ is time,
$c=1/\sqrt{\varepsilon\mu}$ is light speed.

Scalar and vector part of Eq. (37) gives us Hamilton form of Maxwell equations [17,18]:
$$
\rho  = div\,A,\quad \partial _\tau  A + i\,rot\,A + J = 0. \eqno(39)
$$
Real and imaginary part of Eqs.(20) gives us the system of 8 Eqs. of classical form.

If to take mutual bigradient from Eq.(37), we get charge conservation law in scalar part
and wave equation for intensities of EM-field in vector part:
$$\partial _\tau  \rho  + div\,J = 0,\quad  \square A =  i{\kern 1pt} rot\,J -grad\rho  -\partial _\tau  J.$$
Bq. \emph{energy-pulse} of \textbf{A}-field
$${\bf{\Xi}}=W+iP=\frac{1}{2}\,\textbf{A}\circ
\textbf{A}^{\ast}\eqno (40)$$ contains   energy density of EM-field
\[
W = 0,5\left\| A \right\|^2  = \frac{1} {2}\left( {\varepsilon
\left\| E \right\|^2  + \mu \left\| H \right\|^2 } \right) , \]
and Pointing vector
\[
P = \frac{1} {2}[\bar{A} ,A]= c^{ - 1} E \times H .
\]
  \emph{Relativistic formulas } for Maxwell equations for Poincar\'{e} transformation have the form:
$$\nabla^+{\bf A'} = {\bf \Theta'},\quad\quad
\emph{where}     \quad {\bf A}^{\bf '}  = {\bf  P}^- \circ {\bf
A} \circ {\bf P} ,\quad\quad{\bf \Theta}^{\bf '}  = {\bf  P}^-
\circ {\bf \Theta} \circ {\bf P}. $$

\vspace{2mm}

\emph{\textbf{Shock EM-waves.}} From theorem 5.2 we get conditions on fronts of shock EM-waves:
 $$\left[ A\right]_S  = i\bf{m} \circ
\left[ A\right]_S .\eqno (41)$$
Its scalar and vector parts  be of the form :
$$   (\textbf{m},\left[ A\right]_S)=0,\quad
\left[ A\right]_S  =  i[\textbf{m},\left[ A\right]_S ]. $$
First condition shows that shock EM-waves are \emph{transversal}.
From second condition we have
 $$
\sqrt{\varepsilon}\,[ E]_S  = \sqrt{\mu }\, [[ H]_S, \textbf{m}],\quad
\sqrt{\mu }\,\left[H\right]_S  =  \sqrt{\varepsilon}[\textbf{m},\left[ E\right]_S ].$$
Using vectors of electric displacement
and magnetic induction $${D}=\varepsilon E,\quad{B}=\mu H,$$
we state result.\vspace {2mm}

\textbf{Theorem 6.1.} \emph{On the fronts of shock EM-waves  the gaps of intensities of EM-field
satisfy to conditions}:
$
[ E]_S  =c \, [[{B }]_S, \textbf{m}],\quad \left[H\right]_S  = c\,[\textbf{m},\left[D\right]_S ].$
\emph{That is equivalent to conditions:}
$
[ D]_S  = c^{-1} [[H ]_S, \textbf{m}],\quad [B]_S  = c^{-1}[\textbf{m},[E]_S ].$
\vspace {2mm}

\textbf{ \emph{Generalized Kirchhoff formula for Maxwell equation}} be  of the form:
\[
{4\pi }{\textbf{A}} =\nabla^ -  \left\{\int\limits_{r \le \tau
} {{{\bf \Theta}(\tau  - r,y)}}\frac{dV(y)}{r}  +
\tau^{-1}\int\limits_{r =\tau } {\bf A^0}(y) dS(y)\right\} +
\tau^{-1} \int\limits_{r =\tau } {\bf \Theta}(0,y) dS(y).\eqno (42)
 \]
Integral representation for \emph{E,H} can be get from here but they will be  composite form. \vspace{3mm}

\textbf{\emph{Modified Maxwell equations}}.  In Hamilton form of Maxwell Eqs. (34) vector Eq. defines  currents,
scalar Eq. is definition of  a charge.
In bigradient of intensity of EM-field the scalar part is equal to 0. Corollary of it is law of charge conservation.
I.e. Maxwell equations describes closed systems of electric charges and currents and generated by them EM-field.

For open system the Maxwell Eqs. must be modified with use of scalar field $a(\tau,x)$ in Bq.
$\textbf{A}=a(\tau,x)+A(\tau,x)$. Then from Eq. (32) follows its modified form:
\[
 \partial _\tau  A +i\,rotA +J=grad\, a ,\quad \rho  = div\,A -\partial _\tau a.\eqno (43)
\]
If $\rho$ and $J$ are known this system for determination  $a$ and $A$ is closed.
Scalar field $a(\tau,x)$ may be called  \emph{resistance-attraction} field [11,12].\vspace{2mm}

\textbf{\emph{Comment. }}Biquaternionic form of Maxwell equation  (32) has  deep physical sense:
  \emph{electric charges and currents   are physical appearance of EM-field bigradient,
 and intensities of EM-field are the bipotential  for  its charge-current. }
\bigskip

\bigskip

\centerline{{\textbf{7. Biquaternionic presentation of  Dirac operators and their properties
}}}\vspace{2mm}

Biwave equation (20) cam be written in the matrix form:
\[
\sum\limits_{j = 0}^3 {D_{m\,j}^{^ \pm  } } b_j  = g_m ,\quad m,j
= 0,1,2,3,\eqno(44)
\]
 where $b_0  = b,\;g_0  = g,\;b_j  = B_j, \;g_j  = G_j ,\;j = 1,2,3;\;$ and $D_{m{\kern 1pt} j}^ \pm  $ are the
components of matrix $D^ \pm  $ (corresponding to sign), which have the form:
\[
D^ +   = D = \left\{ {\begin{array}{*{20}c}
   {\partial _\tau  } & { - i\partial _1 } & { - i\partial _2 } & { - i\partial _3 }  \\
   {i\partial _1 } & {\partial _\tau  } & { - i\partial _3 } & {i\partial _2 }  \\
   {i\partial _2 } & {i\partial _3 } & {\partial _\tau  } & { - i\partial _1 }  \\
   {i\partial _3 } & { - i\partial _2 } & {i\partial _1 } & {\partial _\tau  }  \\
\end{array}} \right\},\quad D^ -   = \bar D = \left\{ {\begin{array}{*{20}c}
   {\partial _\tau  } & {i\partial _1 } & {i\partial _2 } & {i\partial _3 }  \\
   { - i\partial _1 } & {\partial _\tau  } & {i\partial _3 } & { - i\partial _2 }  \\
   { - i\partial _2 } & { - i\partial _3 } & {\partial _\tau  } & {i\partial _1 }  \\
   { - i\partial _3 } & {i\partial _2 } & { - i\partial _1 } & {\partial _\tau  }  \\
\end{array}} \right\}\eqno(45)
\]
Easy to test that their product (operators superposition)
satisfy to relation:
\[
\sum\limits_{j = 0}^3 {D_{m\,j}^{^{} } } D_{jl}^{^{} }  = \delta
_{m\,l} \Box,\quad j,m,l = 0,1,2,3.\eqno(46)
\]
We'll show that matrixes (45) are the differential matrix operators of Dirac, which possess by such properties
 [13,17]. For this we afford them in matrix form: $$D^+ = \sum\limits_{j = 0}^3 {D^j
\partial _j },\eqno(47)$$ where, as follow from (45), matrixes $D^j $ have the components:
\[
D^0  = I,\;\;D^1  = \left\{ {\begin{array}{*{20}c}
   0 & { - i} & 0 & 0  \\
   i & 0 & 0 & 0  \\
   0 & 0 & 0 & i  \\
   0 & 0 & { - i} & 0  \\
\end{array}} \right\},\quad D^2  = \left\{ {\begin{array}{*{20}c}
   0 & 0 & { - i} & 0  \\
   0 & 0 & 0 & i  \\
   i & 0 & 0 & 0  \\
   0 & { - i} & 0 & 0  \\
\end{array}} \right\},\;D^3  = \left\{ {\begin{array}{*{20}c}
   0 & 0 & 0 & { - i}  \\
   0 & 0 & { - i} & 0  \\
   0 & i & 0 & 0  \\
   i & 0 & 0 & 0  \\
\end{array}} \right\}
\]
here $I$ is unit matrix. As you see, this is 4-dimensional Dirac matrixes, composed from 2-dimensional Pauli matrix:
\[
\left( {\begin{array}{*{20}c}
   1 & 0  \\
   0 & 1  \\
\end{array}} \right),\;\;\left( {\begin{array}{*{20}c}
   { \pm i} & 0  \\
   0 & { \mp i}  \\
\end{array}} \right),\;\;\left( {\begin{array}{*{20}c}
   0 & { \pm i}  \\
   { \pm i} & 0  \\
\end{array}} \right),\;\;\left( {\begin{array}{*{20}c}
   0 & { \pm i}  \\
   { \mp i} & 0  \\
\end{array}} \right).
\]
Therefore  Dirac matrix $D^{\pm}$ has \emph{biquaternionic presentation  $\nabla^{\pm}$ }.

The differential operators:
  \[
  {\bf D}_m^ +   = \nabla ^ +   + m,\quad {\bf D}_m^
-   = \nabla ^ -   + m\,
\]
constitute biquaternionic presentation of matrix Dirac operators: $D^{\pm}+mI$.
By $m=0$ it is  biquaternionic form of Maxwell operator.
By this cause biquaternionic equation of form:
\[
{\bf D}_m^ \pm  {\bf B} \equiv \left( {\nabla ^ \pm   + m} \right)
\circ \,{\bf B} = {\bf F} ,\quad m\in \mathcal{C}, \eqno(48)
\]
is the biquaternionic presentation of generalized  Maxwell-Dirac equation (\emph{MD-equation}).

After simply calculation we get this lemma.

\textbf{Lemma 7.1.}
 \[
{\bf D}_m^ +  \circ{\bf D}_m^ -   = {\bf D}_m^ - \circ {\bf D}_m^ +   = \Box
+ m^2  + 2m\partial _\tau.\]
\emph{For imaginary }$m=i\rho$
\[{\bf D}_{i\rho}^ + \circ {\bf D}_{i\rho}^ -
=  \Box - \rho^2  + 2i\rho\partial
_\tau.\eqno(49)
\]

   It is interesting that right part (49) contains Klein-Gordon-Fokk operator
    ($ \Box- \rho ^2 $), and Sr\'odinger operator ($\Delta + 2i\rho\partial)
_\tau$. By this cause we  will name the equation
\[
\Box u + 2m\partial _\tau  u + m^2 u = f(\tau ,x) \eqno(50)
\]
   \emph{Klein-Gordon-Fokk-Sr\'odinger } equation or shortly
\emph{KGFSh-equation}.

As you'll see further, addition of $2m\partial _\tau$ to KGF-operator essentially simplify
construction of solutions of Maxwell and Dirac equations and their representation.\bigskip

\centerline{\textbf{8.  Generalized solutions of Maxwell-Dirac equation (48).}}
\centerline{\textbf{KGFSh-equation and scalar potentials }}

Using lemma 7.1 and theory of generalized functions it is easy
 to construct generalized solutions of MD-equation(48).

\textbf{Theorem  8.1.} \emph{Generalized solutions  of MD-equation} (48) \emph{have the form}:
\[
{\bf B} = {\bf B}^0  + {\bf D}_m^ \mp \circ \left( {\psi  * }
{\bf F} \right)= {\bf B}^0  + {\bf{\Psi}}\ast{\bf F},\eqno(51)
\]
\emph{where} ${\bf B}^0 (\tau ,x)$ is  \emph{a solution of homogeneous equation}:
\[
{\bf D}_m^ \pm  \circ{\bf B}^0  = 0,\eqno(52)
\]
${\bf{\Psi}}$ \emph{is its fundamental solution}:
$${\bf{\Psi}}(\tau,x)=\partial_\tau\psi+m\psi\pm i\, grad\,\psi,$$
$\psi (\tau ,x)$ is \emph{fundamental solution of KGFSh-equation}(50).
\vspace{2mm}

\textbf{Proof. }In virtue of linearity of equation, it is sufficiently to prove this assertion for
 the second summand in  formula (51).
If to substitute it to Eq.(48), by use of properties (15) and $\delta$-function, we get:
   \[
{\bf D}_m^ \pm  {\bf D}_m^ \mp  \,\left( {\psi  * } \right.\left.
{\bf F} \right) = {\bf F} * \left( {\Box \psi  + 2m\partial _\tau
\psi + m^2 \psi } \right) = {\bf F} * \delta (\tau )\delta (x) =
{\bf F}.
\]
Ex facte, whatever solution of Eq. (48) can be presented in form (51).\vspace{2mm}

 Let formulate this theorem for imaginary  $m =
i\rho $ .\vspace{2mm}

    \textbf{Theorem 8.2.} \emph{Generalized solutions of MD-equation}     :
$$\left( {\nabla ^ \pm   + i\rho } \right)\circ{\bf B} = {\bf F} \eqno(53)$$
\emph{ can be presented in the form:
\[
 {\bf B} = {\bf B}_0  + \left( {\nabla ^ \mp   +
i\rho } \right) \circ\left( {{\bf F} * }  {\psi
 } \right),\eqno(54)
\]
where
$\psi  (\tau ,x) $ is fundamental solution of KGFSh-equation}:
\[
\Box \psi    - \rho ^2 \psi    + 2i\rho \,\partial _\tau \psi
   = \delta (\tau )\delta (x),\eqno(55)
\]
${\bf B}_0 (\tau ,x)$ \emph{ is a solution of Dirac equation}:
\[
\left( {\nabla ^ \pm   + i\rho } \right)\circ{\bf B}_0  = 0.\eqno(56)
\]

From these theorems it follows  that  solutions of KGFSh-equation determine  solutions of MD-equation.
\vspace{2mm}

\textbf{Theorem 8.3  }  \emph{Generalized solutions  of KGFSh-equation (50) are presented in the form: $u = f *
\psi + u_0$,  where $\psi$ is its fundamental solutions}:
 \[ \psi  = \frac{1}{{4\pi
\left\| x \right\|}}\left( {a\,e^{ - m\left\| x \right\|}
\delta (\tau  - \left\| x \right\|) + (1 - a)\delta
(\tau  + \left\| x \right\|)e^{m\left\| x \right\|} } \right) +
\psi _0 ,\quad \forall a\in \mathcal{C},
\]
$\delta (\tau  \pm\left\| x \right\|)$ \emph{is simple fiber on light cone }$\left\| x
\right\|=\mp \tau $ , $u_0 (\tau ,x)$ \emph{is a solution of homogeneous Eq.,
which exists only for imaginary $m=i\rho$  and}
\[
u_0  (\tau ,x) = e^{ - i\rho \tau } \int\limits_{R^3 } {\phi
(\xi )} \exp \left( {i\left( {(\xi ,x) \pm \left\| \xi
\right\|\tau } \right)} \right)dV(\xi ) ,\quad  \forall \phi (\xi ) \in L_1 (R^3 ).\eqno (57 )
\]
\vspace{2mm}

 \textbf{Proof. }From equation for fundamental solution:
    \[
\Box \psi  + m^2 \psi  + 2m\partial _\tau  \psi  = \delta (\tau
)\delta (x),\eqno (58 )
\]
follows Helmholtz equation:
\[
\left\{ {\Delta  - k^2 } \right\}F_\tau  [\psi ] + \delta (x) =
0,\quad k = i\omega  - m.
\]
fundamental solutions of which are well known:
\[
F_\tau  [\psi  ] = \frac{1}{{4\pi \left\| x \right\|}}\left(
{ae^{(i\omega  - m)\left\| x \right\|}  + (1 - a)e^{ - (i\omega  -
m)\left\| x \right\|} } \right) ,\quad \forall a\in \mathcal{C}. \]
 From here by use of properties of direct and inverse Fourier transformation the first formula of  theorem follows.

Support of first summand is expanding over time  sphere of radius $ \tau$ ( $\tau
> 0$), and for second one it is converging over time sphere of radius $\left| \tau
\right|$ ($\tau  < 0$).

If $m  $ is imaginary ( $m = i\rho $ ) and support of solution over time is $\tau  >
0$, then
\[
\psi  = \frac{{e^{ - i\rho \left\| x \right\|} }}{{4\pi \left\|
x \right\|}}\delta (\tau  - \left\| x \right\|). \eqno(59)\]
It's interesting that here density of simple fiber in light cone is fundamental solutions of Helmholtz Eq. with wave number
$\rho $. \vspace{2mm}

\textbf{\emph{Solutions of homogeneous KGFSh-equation}}. Let consider
\[
\Box u + m^2 u + 2m\partial _\tau  u = 0 \eqno(60)
 \]
which in Fourier transform space satisfy to algebraic Eq.
\[
 \left({\left\| \xi  \right\|^2  - (\omega  + im)^2 } \right)u^* (\omega
,\xi ) = 0.\eqno(61)
 \]
If ${\mathop{\rm Re}\nolimits} \,m \ne 0$, then $\left\| \xi
\right\|^2  - (\omega  + im)^2  \ne 0$ by $\forall\{ \xi,\omega\}  \in R^4$.
In this case the equation have only trivial solution: $u^*  = 0$.

Provided  $m = i\rho $
this Eq. has infinitely set of singular solutions of the type:
\[
u^* (\omega ,\xi ) = \phi (\omega ,\xi )\delta \left( {\left\| \xi
\right\|^2  - (\omega  - \rho )^2 } \right) , \eqno(62)\]
 where
$\phi (\omega ,\xi )$ is a density of simple fiber. It is arbitrary given and integrable   function on cones
 $\left\| \xi  \right\| = \left|
{\omega  - \rho } \right|$ .  By calculating of originals, we get:
\[
u(\tau ,x) = \frac{1}{{\left( {2\pi } \right)^4 }}\int\limits_{ -
\infty }^\infty  {d\omega } \int\limits_{\left\| \xi  \right\| =
\left| {\omega  - \rho } \right|} {\phi (\omega ,\xi )} \exp
\left( { - i(\xi ,x) - i\omega \tau } \right)dS(\xi ) =
\]
\[
 = \frac{{e^{ - i\rho \tau } }}{{\left( {2\pi } \right)^4 }}\int\limits_{R^3 }
  {\left\{ {\phi (\rho  + \left\| \xi  \right\|,\xi )e^{ - i\left\|
   \xi  \right\|\tau }  - \phi (\rho  - \left\| \xi  \right\|,\,\,
   \xi )e^{i\left\| \xi  \right\|\tau } } \right\}} \exp \left( { - i(\xi ,x)} \right)d\xi
_1 d\xi _2 d\xi _3.
\]
In virtue of arbitrary  $\phi$, we get formula (57). The theorem is proved.\vspace{2mm}

\textbf{ \emph{Harmonic vibrations}}. Let construct solutions of MD-equation (48) in case
of  harmonic vibration with frequency $\omega$:
$${\bf B} = {\bf B}(x)e^{  i\omega \tau },\quad{\bf F} = {\bf F}(x)e^{  i\omega \tau }.$$
Then  we have the equation for complex amplitude of type :
\[
\nabla _{( \omega   + \rho )}^{ \pm  } \,{\bf B}(x) \equiv( \omega   + \rho +\nabla )\circ{\bf B}(x)=
{\bf F}(x) . \eqno (61) \]
which be named \emph{gradiental}.
It easy to calculate
$$\nabla _{( \omega )}\circ \nabla _{( -\omega )}=-(\omega^2+\bigtriangleup). $$
\vspace{2mm}
By using this property it easy to prove the theorem.\vspace{2mm}

\textbf{Theorem 8.4. }\emph{Generalized solutions of gradiental equation} (61) \emph{can be formulary}:
\[
{\bf B} =  {\nabla _{\omega + \rho }^ \mp   } \left( {\chi
* } \right.\left. {\bf F} \right) + {\bf Sp}^{(\omega+\rho)},
\]
\emph{where }  $$ \chi  =  - \frac{{ae^{-ik\left\| x \right\|}
}}{{4\pi \left\| x \right\|}} + \frac{{(a-1)e^{  ik\left\| x
\right\|} }}{{4\pi \left\| x \right\|}},\quad k = \left| {\omega +
\rho } \right| \ne 0,\quad \forall a\in \mathcal{C},\quad \eqno(62)$$
$$ {\bf Sp}^{(\omega+\rho)} = {\nabla _{\omega+ \rho } ^ \mp   }\left( {\chi _0
* {\bf C}(x)} \right),$$
$\chi _0 (x)$ \emph{is a solution of Helmholtz equation with wave number $k$}:
\[ \chi _0 (x) =
\int\limits_{\left\| {\bf e} \right\| = 1} {p({\bf e})} \;e^{ -
ik({\bf e},x)} dS({\bf e}),
\]
\emph{$p({\bf e})$ is  $\forall $ integrable on unit sphere function,
${\bf C}(x)$ is $\forall$ Bq., for which this convolution exists. }\vspace{1mm}

If $\omega>0$ first summand in (62) describes radiated  spherical wave, second one -- converging spherical wave.
The wave number $k$ defines the length of these waves $\lambda=2\pi/k$, which
(for positive $\rho$) decreases by increasing $\rho$. But if $\rho<0$ then $\lambda$ in interval
$\{-\omega,0\}$ increases by increasing $|\rho|$ and by $\rho=-\omega$ spherical waves vanished.\vspace{2mm}

\emph{\textbf{Statics.}}This theorem also defines  solutions of static MD-equation (when $\omega=0$) .
\vspace{2mm}

\textbf{Theorem 8.5. }\emph{Generalized solutions of static MD-equation  can be formulary}:
\[
{\bf B} =  {\nabla _{\omega }^ \mp   } \left( {\chi
* }  {\bf F} \right) + {\bf Sp}^{(\rho)},
\]
\emph{where }  $$ \chi  =  - \frac{{ae^{-i\left\| \rho x \right\|}
}}{{4\pi \left\| x \right\|}} + \frac{{(a-1)e^{  i\left\|\rho x
\right\|} }}{{4\pi \left\| x \right\|}},
 \quad \forall a\in \mathcal{C},$$
$$ {\bf Sp}^{(\rho)} = {\nabla _{\rho } ^ \mp   }\left( {\chi _0
* {\bf C}(x)} \right),$$
$\chi _0 (x)$ \emph{is a solution of of Helmholtz equation with wave number $|\rho|$}:
\[ \chi _0 (x) =
\int\limits_{\left\| {\bf e} \right\| = 1} {p({\bf e})} \;e^{ -
i|\rho|({\bf e},x)} dS({\bf e}),
\]
\emph{$p({\bf e})$ is  $\forall $ integrable on unit sphere function,
${\bf C}(x)$ is $\forall$ Bq., for which this convolution exists. }\vspace{2mm}

We remark that Bq. $\textbf{C}\in \textrm{B}'(M)$. For regular \textbf{B} this formulas gives classic solutions of
MD-equations.

If $\rho=0$ from these two theorems we get solutions of modified Maxwell equations in cases of harmonic vibrations
and statics.

\bigskip

\centerline{{\textbf{9. Spinors and spinors fields} }} \vspace{2mm} In quantum mechanics solutions of \emph{Dirac equation}:
\[
 \left( {\nabla ^ \pm   + i\rho} \right)\circ{\bf Sp} = 0,\quad
\textrm{ Re}\rho=0.    \eqno (63)
 \]
are called   \emph{spinors}  [13,17,19].\vspace{2mm}

\textbf{ Theorem  9.1.} \emph{Spinors may be formulary}:
\[
{\bf Sp} = {\bf D}_{i\rho}^ \mp  \circ\left( {\psi _0  * {\bf C}(\tau
,x)} \right) ,      \eqno (64)
\]
 \emph{where  $\psi _0 $ is  a solution of homogeneous KGFSh-equation},  \emph{ or can be  presented as the
sum of such form solutions, $\forall{\bf C}\in {{\texttt{B}'}}{(M)}$ for which this convolution exists.}\vspace{2mm}

\textbf{Proof.}  By substituting formula (64) into Eq.(63)
we get:
\[
{\bf D}_m^ \pm  {\bf Sp} = {\bf D}_m^ \pm  {\bf D}_m^ \mp  \left(
{\psi _0 * {\bf C}} \right) = \left( \Box{\psi _0 + 2m\partial
_\tau  \psi _0 + m^2 \psi _0} \right) * {\bf C} = 0,\quad m=i\rho.
\]
Inversely, if spinor (64) is a solution of (63), then
\[
\left( { \Box+ 2m\partial _\tau   + m^2 } \right)\circ{\bf Sp} = {\bf
D}_m^ \mp  {\bf D}_m^ \pm  {\bf Sp} = {\bf D}_m^ \mp  0 = 0.
\]
Therefore scalar part and components of vector part ${\bf Sp}$
are the solutions of KGFSh-equation and
${\bf Sp}$   may be formulary as the sum of solutions of type (64).

   In particular case, when ${\bf C} = 1$, we have \emph{spinor of scalar $\psi _0$-field } :
\[
{\bf \Psi }_0^ \mp  = \left( {\nabla ^ \mp   + m} \right)\circ\psi _0 =
m\psi _0  + \partial _\tau  \psi _0  \mp i\,\textrm{grad}\psi _0,
\]
Consequently \textbf{C}-\emph{field spinors }, generated by potential $\psi _0$,
 may by formulary:
\[
{\bf Sp}^{\pm} = {\bf \Psi }_0^{\pm}   * {\bf C}(\tau ,x).     \eqno (65)
\]

\textbf{\emph{{ Scalar harmonic potentials.}}}
Let see formula (57) where there are two plane harmonic wave:
\[
\varphi _\xi ^ \pm  (\tau ,x) = \exp \left( {i\left( {(\xi ,x) -
\rho \tau  \pm \left\| \xi  \right\|\tau } \right)} \right).
\]
They also are solutions of  homogeneous KGFSh-equation. Wave vector $\xi $
defines direction of wave motion. The length of these waves is equal to
$\lambda  = 2\pi /\left\| \xi \right\|$, their frequencies are $\omega  = \left|
{\rho  \pm \left\| \xi \right\|} \right|$, periods $T = 2\pi
/\left| {\rho  \pm \left\| \xi \right\|} \right|$.
In depend of sign, one of them has the supersonic phase speed ($V>1$) ,
other one has the subsonic phase speed   ($V<1$) , as  $V = 1 \pm \frac{\rho }{{\left\| \xi  \right\|}}$.
$V \to 1 \pm 0$  by
$\left\| \xi \right\| \to \infty $ ,
but $\omega  \to \infty $. By $\left\| \xi \right\| \to \left| \rho
\right|$ velocities $V \to 1;0$ and frequencies  $\omega
\to \frac{\pi }{\rho };\;\infty $ (corresponding to sign).

Generated by these waves spinors
\[
\left( {\nabla ^ \mp   + i\rho } \right)\circ\varphi _\xi ^ \pm  (\tau
,x) =  \pm \left( {i\left\| \xi  \right\| + \xi } \right)\varphi
_\xi ^ \pm .
\]
have the form, norm and pseudonorm:
\[ {\bf Sp}_\xi ^ \pm   = \frac{\exp \left( {i\left( {(\xi ,x) -
\rho \tau  \pm \left\| \xi  \right\|\tau } \right)}
\right)}{{\sqrt 2 }}\left( {i + \frac{\xi }{{\left\| \xi
\right\|}}} \right) ,\] \[ \left\| {{\bf Sp}_\xi ^ \pm  } \right\|
= 1 ,\quad \left\langle\langle {{\bf Sp}_\xi ^ \pm  } \right\rangle  \rangle =
0.\]
We name them \emph{elementary $\xi$-oriented harmonic spinors}.
 Its energy-impulse is equal to
\[
{\bf \Xi } = {\bf Sp}_\xi ^ \pm   \circ ({\bf Sp}_\xi ^ \pm  )^ *   = 1-i\frac{\xi}{\|\xi\|} \,\Rightarrow\,
\left\| {\bf \Xi } \right\|
= 2 ,\quad \left\langle\langle {\bf \Xi }  \right\rangle  \rangle =0.
\]
\emph{{${\bf C}$-field  $\xi$-oriented harmonic spinor  }} is the spinor:
\[
{\bf Cp_{\xi}} = {\bf C}(\tau ,x) * {\bf Sp}_\xi ^ \pm  (\tau ,x).\eqno
(66)
 \]
From formula (64) it follows that
${\bf C}$-field harmonic  spinors  may be formulary as
\[
{\bf Cp} = {\bf C}(\tau ,x) * \int\limits_{R^3 } {{\bf
Sp}_\xi ^ \pm  (\tau ,x)}\phi (\xi ) d\xi_1d\xi_2d\xi_3,\quad \phi (\xi ) \in L_1 (R^3 ),
\]
{{\emph{Elementary harmonic $(\omega+\rho)$-spinors }}}  are defined as:
\[
{\bf \Psi }_0^{(\omega+\rho)}  (x,{\bf e}) = \frac{1}{{k\sqrt 2
}}\left( {\nabla   +\omega  + \rho } \right)\circ e^{ - ik({\bf
e},x)}  =  \frac{1}{{k\sqrt 2 }}\left( {\omega  + \rho  - ik{\bf
e}} \right)e^{ - ik({\bf e},x)} \eqno (68)
\]
It easy to test that
\[
\left\| {{\bf \Psi }_0^{\omega +\rho} } \right\| = 1,\quad
\left\langle \left\langle {{\bf \Psi }_0^{\omega +\rho}  } \right\rangle\right\rangle   = 0.\eqno (69)
                                              \]
Here  ${\bf e}$ is the direction of the spinor, $k = \left|
{\omega  + \rho } \right|$ is its wave number.
Energy-impulse of ${{\bf \Psi }_0^{(\omega +\rho)}  }$ are equal to
\[
{\bf \Xi } = {\bf \Psi }_0^{(\omega+\rho)}  \circ \{{\bf \Psi }_0^{(\omega+\rho)}\}^ *   = 1- i\,\textbf{e}\,sign(\omega +\rho)
\]

\textbf{Theorem 9.1. }  \emph{  {\bf G}-field harmonic spinors can be formulary}
\[
{\bf Gp}^{\omega+\rho} (x,e) = {\bf G}(x) * {\bf \Psi }_0^
{\omega+\rho} (x,{\bf e})\quad\quad (\emph{e-oriented spinors field})\]
\emph{or}
\[
{\bf Gp}^{\omega+\rho} (x) = {\bf G}(x) * \int\limits_{\left\| {\bf
e} \right\| = 1} {p({\bf e}){\bf \Psi }_0^{\omega+\rho}  (x,{\bf
e})} dS({\bf e})\quad (\emph{nonoriented spinors field})\]
  $ \forall p({\bf e}) \in L_1
(\{ {\bf e} \in R^3 :\left\| {\bf e}
\right\| = 1\} )$.\vspace{2mm}

\textbf{\emph{ Static spinors}} are obtained by $\omega  = 0$.

\bigskip

{\textbf{Conclusion.}}The solutions  of  biquaternionic form of Maxwell-Dirac  equations
 are here received in class of generalized function. That allows to
 build the decisions as for regular biquaternionic functions, so
 and at presence of singular sources in its right part. It's possible to use at
  building of biquaternionic theories of the elementary particles.  At calculation
   of spinors fields,  it is possible
   to flip differentiation on components \textbf{\emph{C}}-field, when this suitable.
\textbf{\emph{C}}-field too can be singular generalized function.

Bigradients, biwave equations and their decisions were used by author earlier for
 building of one models of electro-gravymagnetic fields and interactions [11,12,20].
It's possible to find much other useful applications of the differential algebra
of biquaternions, what is offered  to interested reader.\vspace{5mm}

\textbf{Key words}: \emph{biquaternion, bigradient, biwave equation,
Maxwell-Dirac equations, generalized solutions, shock waves, spinors,
harmonic spinors, spinors field }
\vspace{3mm}

 \centerline{  \textbf{References}}

{1.} W.R. Hamilton. \emph{ On a new Species of Imaginary Quantities connected
with a theory of Quaternions}, Proceedings of the Royal Irish
Academy, \textbf{2} (Nov 13,1843), 424-434.

{2. }J.D. Edmonds Jr.\emph{Eight Maxwell equations as one quaternionic}, Amer. J. Phys.,  \textbf{46 }(1978), No. 4, 430

{3. } G.L. Shpilker.\emph{Hypercomplex solutions of Maxwell equations}, Report of USSR Academy of Sciences  ,  \textbf{272} (1983), № 6, 1359-1363

{4. } M. Acevedo M., J. Lopez-Bonilla and M. Sanchez-Meraz. \emph{Quaternions, Maxwell
Equations and Lorentz Transformations},  Apeiron, \textbf{ 12} (2005), No. 4 , 371

{5. } S.L. Adler. \emph{Quaternionic quantum mechanics and quantum fields}, New York: Oxford University Press,1995.

{6. } P. Rotelli. \emph{The Dirac equation on the quaternionic field}, Mod. Phys. Lett. A 4 (1989), 933-940 .

{7. }A.J. Davies,  \emph{Quaternionic Dirac equation}, Phys. Rev. D \textbf{41}(1990), 2628-2630 .

{8. } D. Finkelstein , J. M. Jauch , S. Schiminovich, D. Speiser.
 \emph{Foundations of quaternion quantum mechanics}, J. Math. Phys., \textbf{3}(1992), 207-220 .

{9. } S. De Leo, W. A. Rodrigues Jr.  \emph{Quaternionic electron theory:
 geometry, algebra and Dirac's spinors}, Int. J. Theor. Phys., \textbf{37}(1998), 1707-1720 .

{10. }  V.V. Kassandrov.  \emph{Biquaternion
electrodynamics and Weyl-Cartan    geometry of space-time},    Gravitation and cosmology, \textbf{1} (1995), 3, 216-222.

{11.}  L.A.Alexeyeva. \textit{Equations of interaction of A-fields and Newton laws},
Intelligence of National Academy of Sciences of Rep.Kazakhstan, Physical and mathematical series , No.3 (2004), 45-53.

{12. } V. V. Kravchenko. \emph{On force-free magnetic fields: quaternionic approach}. Mathematical Methods
in the Applied Sciences, 2005, v. 28, No. 4, 379-386.

{13. }N.V.Bogoljubov, A.A.Logunov, A.A.Oksak, I.T.Todorov. \emph{General principles to quantum theory of the field.}
 Moscow:Science, 1987. 616 p.

{ 14.}  L.A.Alexeyeva.\emph{Differential algebra of biquaternions.} 1.\emph{ Lorentz transformations},
Mathe\-matical journal, \textbf{10} (2010),No.1, 33-41.

{15. } V.S.Vladimirov. \textit{Generalized functions in mathematical physics},
Moscow:Science, 1976, 512 p.

{16. }   L.A.Alexeyeva. \textit{Quaternions of Hamilton form of Maxwell equations}, Mathe\-matical journal,
\textbf{3}(2003), No.4,20-24.

{17.} A.I.Ahiezer, D.B.Berestezkij. \emph{Quantum electrodynamics}, Moscow:Science, 1981. 320 p.

{18.}   L.A.Alexeyeva. \emph{Hamilton form
 of Maxwell equations and its generalized solutions},
Differential equations,\textbf{39} (2003), No.6, 769-776.

{19.} \emph{Mathematical encyclopedia},
Moscow:Science,\textbf{2} (1982).

{20} Alexeyeva L.A.\emph{ Newton's laws for a biquaternionic model of the electro-gravimagnetic fields,
charges, currents, and their interactions}, Ashdin publishing.
Journal of Physical Mathematics. 2009. Vol.1.  Article ID  S090604

\vspace{2mm}
 Paper was submitted    ----------------   2011 г.
\bigskip

Alexeyeva Luydmila Alexeyevna,\\ Professor, Doctor of physical and mathematical sciences,\\
Head of Waves dynamics laboratory,\\
Institute of Mathematics,\\
Ministry of Education and Sciences,\\
Kazakhstan.\\

Address:
Pushkin str. 125, Almaty, Kazakhstan,  050010

Telephone +77773381814

E-mail: alexeeva@math.kz, alexeeva47@mail.ru

\end{document}